\documentclass[%
 reprint,
superscriptaddress,
 amsmath,amssymb,
 aps,
 prl,
]{revtex4-1}

\usepackage{graphicx}
\usepackage{dcolumn}
\usepackage{bm}
\usepackage{natbib}
\usepackage{hyperref}
\usepackage{siunitx}
\usepackage[caption=false]{subfig}
\usepackage{booktabs}
\usepackage[export]{adjustbox}
\usepackage{multirow}
\usepackage{physics}
\usepackage{accents}

\usepackage{color}
\usepackage{multibib}
\usepackage{xspace}

\begin{document}

\renewcommand{\m}[1]{\ensuremath{\mathbf{#1}}}
\renewcommand{\v}[1]{\ensuremath{\mathbf{#1}}}
\newcommand{\expens}[1]{\mathbb{E}[\langle {#1} \rangle]}
\newcommand{\expq}[1]{\langle {#1} \rangle}
\newcommand{\exps}[1]{\mathbb{E}[ {#1} ]}

\def\psihat{\ensuremath{\hat{\psi}}\xspace}
\def\rhochat{\ensuremath{\hat{\rho}_c}\xspace}
\def\psihatd{\ensuremath{\hat{\psi}^{\dagger}}\xspace}
\def\ahat{\ensuremath{\hat{a}}\xspace}
\def\Ham{\ensuremath{\hat{H}}\xspace}
\def\Innov{\ensuremath{\mathcal{H}}\xspace}
\def\ahatd{\ensuremath{\hat{a}^{\dagger}}\xspace}
\def\bhat{\ensuremath{\hat{b}}\xspace}
\def\bhatd{\ensuremath{\hat{b}^{\dagger}}\xspace}
\def\boldr{\ensuremath{\mathbf{r}}\xspace}
\def\dr{\ensuremath{\,d^3\mathbf{r}}\xspace}
\def\dk{\ensuremath{\,d^3\mathbf{k}}\xspace}
\def\etal{\emph{et al.\/}\xspace}
\def\ie{i.e.\:}
\def\Qhat{\ensuremath{\hat{Q}}\xspace}
\def\Qhatd{\ensuremath{\hat{Q}^\dag}\xspace}
\def\phihatd{\ensuremath{\hat{\phi}^{\dagger}}\xspace}
\def\phihat{\ensuremath{\hat{\phi}}\xspace}
\def\boldk{\ensuremath{\mathbf{k}}\xspace}
\def\boldp{\ensuremath{\mathbf{p}}\xspace}
\def\boldsigma{\ensuremath{\boldsymbol\sigma}\xspace}
\def\boldalpha{\ensuremath{\boldsymbol\alpha}\xspace}

\def\Psihat{\ensuremath{\hat{\Psi}}\xspace}
\def\Psihatd{\ensuremath{\hat{\Psi}^{\dagger}}\xspace}
\def\Vhatd{\ensuremath{\hat{V}^{\dag\mathcal{Q}ger}}\xspace}
\def\Xhat{\ensuremath{\hat{X}}\xspace}
\def\Xhatd{\ensuremath{\hat{X}^{\dag}}\xspace}
\def\Yhat{\ensuremath{\hat{Y}}\xspace}
\def\Yhatd{\ensuremath{\hat{Y}^{\dag}}
\xspace}
\def\jhat{\ensuremath{\hat{J}}
\xspace}
\def\lhat{\ensuremath{\hat{L}}
\xspace}
\def\Nhat{\ensuremath{\hat{N}}
\xspace}
\def\ddt{\ensuremath{\frac{d}{dt}}
\xspace}
\def\nset{\ensuremath{n_1, n_2,\dots, n_k}
\xspace}


\title{A new approach to the quantization of the damped harmonic oscillator}

\author{Matthew J. Blacker}
\affiliation{Department of Quantum Science, Research School of Physics, 
Australian National University, Canberra 2601, Australia}%
\author{David L. Tilbrook}
\email{david.tilbrook@anu.edu.au}
\affiliation{Department of Theoretical Physics, Research School of Physics, Australian National University, Canberra 2601, Australia}%

\date{\today}

\begin{abstract}
In this paper, a new approach for constructing Lagrangians for driven and undriven linearly damped systems is proposed, by introducing a redefined time coordinate and an associated coordinate transformation to ensure that the resulting Lagrangian satisfies the Helmholtz conditions. The approach is applied to canonically quantize the damped harmonic oscillator and although it predicts an energy spectrum that decays at the same rate to previous models, unlike those approaches it recovers the classical critical damping condition, which determines transitions between energy eigenstates, and is therefore consistent with the correspondence principle. It is also demonstrated how to apply the procedure to a driven damped harmonic oscillator.
\end{abstract}


\maketitle


\section{\label{sec:introduction}Introduction}
Harmonic oscillators experiencing linear and non-linear damping, both with and without a driving force, arise in a range of physical contexts including modelling superconducting qubits using Josephson junctions \cite{kjaergaard_superconducting_2020,devoret_superconducting_2013,mitra_quantum_2008}, heavy ion scattering \cite{iwasa_magnetic_2019,fujii_communication_2013,ghosh_coherent_1981} and nuclear fission \cite{verriere_improvements_2021,isar_damped_2006,wu_quantum_1988}. There is consensus in the literature that quantization techniques for the linearly damped, undriven harmonic oscillator 
\begin{align}
    \ddot{q} + 2 \alpha \dot{q} + \omega^2 q = 0,
    \label{DampedHarmonicOscillatorEquation}
\end{align}
can be generalised to the above contexts \cite{um_quantum_1987,caldeira_quantum_1983}. Here, $q$, $\alpha$ and $\omega$ are the position coordinate, the coefficient of linear damping and oscillator frequency respectively.\\\\
Existing canonical quantization procedures, however, do not work for non-conservative systems, i.e. it has not been possible with approaches previously reported in the literature to write a Lagrangian which produces equation \eqref{DampedHarmonicOscillatorEquation} as its Euler-Lagrange equation, as it does not satisfy the Helmholtz conditions \cite{engels_helmholtz_1975}. The attempts to overcome this problem have resulted in a wide range of proposals over the last century, the most prominent being that of Bateman \cite{bateman_dissipative_1931}, Caldirola \cite{caldirola_forze_1941}, and Kanai \cite{kanai_quantization_1948} (hereafter referred to as the BCK approach). In the BCK approach, one quantises the Bateman Lagrangian $\mathcal{L}_B$
\begin{align}
    \mathcal{L}_B = \frac{1}{2} \left( \dot{q}^2 - \omega^2 q^2 \right) e^{2\alpha t},
    \label{BatemanLagrangian}
\end{align}
from which the Euler-Lagrange equations produces equation \eqref{DampedHarmonicOscillatorEquation} multiplied by the integrating factor $e^{2\alpha t}$. The canonical momentum associated with $q$ is $p = e^{2\alpha t} \dot{q}$, and can be used to write the corresponding Hamiltonian $H_{BCK}$
\begin{align}
    H_{BCK} = \frac{1}{2} \left( e^{-2\alpha t} p^2 + \omega^2 e^{2\alpha t} q^2 \right).
    \label{BCKHamiltonian}
\end{align}
Many other proposed approaches, such as those employing separation of variables or redefinition of the position variable $q$, are equivalent to this BCK approach \cite{um_quantum_1987,feshbach_quantization_1977,banerjee_canonical_2002,gitman_action_2007,serhan_quantization_2018,rabei_quantisation_2019}. Another class of approaches involve coupling an undamped oscillator to a loss mechanism \cite{senitzky_dissipation_1960,caldeira_quantum_1983,ford_statistical_1965,dekker_quantization_1977}, including through a spin-boson model \cite{thorwart_dynamics_2004,wilhelm_spin-boson_2004} or a Linbladian master equation formalism \cite{isar_density_1993,isar_uncertainty_1999,fujii_quantum_2013}, however these approaches require a model of how the dissipation occurs, i.e. they do not provide a self-contained description of the phenomena associated with a damped harmonic oscillator within a closed system. The BCK approach therefore remains as the most well-known attempt to quantize the damped harmonic oscillator, and solutions to $H_{BCK}$ have been thoroughly investigated \cite{um_quantum_1987,greenberger_critique_1979,hasse_quantum_1975}.\\\\
There are two features of the BCK approach which have recently caused debate in the literature. Firstly, it was proved that a square-integrable vacuum cannot be found for the BCK Hamiltonian \cite{bagarello_no-go_2019}, a result which has withstood some debate \cite{deguchi_square-integrable_2019,bagarello_reply_2019,bagarello_remarks_2020}. Secondly, in Ref. \cite{deguchi_quantization_2020} the authors claim that equation \eqref{BatemanLagrangian} describes a doubled system of both a damped harmonic oscillator and an amplified harmonic oscillator (equation \eqref{DampedHarmonicOscillatorEquation} with $\alpha < 0$). In that paper, they propose a modified Bateman Lagrangian, by introducing additional real dynamical variables and obtain a ladder of energies
\begin{align}
    E_n(t) = \hbar \omega e^{-2\alpha t} \left( n + \frac{1}{2} \right),
    \label{DeguchiEnergySpectrum}
\end{align}
and find a critical damping condition for the quantized system, $\alpha = \omega ( \sqrt{5} - 1)/2$, which differs from the classical condition $\alpha = \omega$. Their approach, however, does not satisfy the correspondence principle and relies upon modifying an existing Lagrangian with multiple dynamical variables. \\\\
In this work, we propose a new approach for quantizing driven and undriven linearly damped systems of the form
\begin{align}
    \ddot{q} + 2\alpha \dot{q} + \sum_i g_i(q,t) = 0,
        \label{GeneralDrivenOscillatorEquation}
\end{align}
where $g_i(q,t)$ is an arbitrary continuous function of $q$ and $t$. Our approach only requires a single dynamical variable, does not require modification of an existing Lagrangian and is therefore easily generalized. We do so by introducing a new time coordinate in equation \eqref{GeneralDrivenOscillatorEquation}, for which we find an exact Lagrangian and Hamiltonian which is then quantized. Importantly, whilst we predict the energy spectrum given by equation \eqref{DeguchiEnergySpectrum}, we recover the classical critical damping condition $\alpha = \omega$. Therefore, our quantization satisfies the correspondence principle. Finally, we demonstrate how to apply our approach to driven linearly damped systems, by considering an example from superconducting quantum computing.
\section{Deriving a Hamiltonian via the Helmholtz Conditions}
For a set of coordinates $q_i$ and their derivatives with respect to time $t$, the Helmholtz conditions state that a Lagrangian can be written for a system of differential equations $E_i \left( t, q_i, \dot{q}_i, \ddot{q}_i \right)$ if they satisfy \cite{engels_helmholtz_1975}
\begin{subequations}
\begin{align}
    \frac{\partial E_i}{\partial \ddot{q}_k} - \frac{\partial E_k}{\partial \ddot{q}_i} &= 0, \label{FirstHelmholtzCondition} \\
    \frac{\partial E_i}{\partial \dot{q}_k} + \frac{\partial E_k}{\partial \dot{q}_i} - \frac{d}{dt} \left( \frac{\partial E_i}{\partial \ddot{q}_k} + \frac{\partial E_k}{\partial \ddot{q}_i} \right) & = 0, \label{SecondHelmholtzCondition} \\
    \frac{\partial E_i}{\partial {q}_k} - \frac{\partial E_k}{\partial {q}_i} - \frac{1}{2} \frac{d}{dt} \left( \frac{\partial E_i}{\partial \dot{q}_k} - \frac{\partial E_k}{\partial \dot{q}_i} \right) & = 0. \label{ThirdHelmholtzCondition}
\end{align}
\end{subequations}
A single differential equation trivially satisfies equations \eqref{FirstHelmholtzCondition} and \eqref{ThirdHelmholtzCondition}; however, equation \eqref{GeneralDrivenOscillatorEquation} (and thus equation \eqref{DampedHarmonicOscillatorEquation}) does not satisfy equation \eqref{SecondHelmholtzCondition}. Therefore, in this coordinate system, it is not possible to write a Lagrangian which produces equation \eqref{DampedHarmonicOscillatorEquation} as its Euler-Lagrange equation. \\\\
We propose considering a new time coordinate $\tau = f(t)$, where $f(t)$ is a continuous and differentiable function for all $t \geq 0$, such that the Helmholtz conditions are satisfied. Equation \eqref{GeneralDrivenOscillatorEquation} becomes
\begin{align}
\begin{split}
    \left( \frac{d\tau}{dt} \right)^2 \ddot{q}\left(\tau\right) & +\left[ \left( \frac{d^2 \tau}{dt^2} \right) + 2 \alpha \left( \frac{d\tau}{dt} \right) \right] \dot{q}\left( \tau \right) \\
    & + \sum_i f_i \left( q \left( \tau \right), \tau \right) = 0,
    \end{split}
    \label{EquationOfMotionRedefinedTimeCo}
\end{align}
assuming that $d\tau/dt \not =0$ for all times $t$ of interest, that $\alpha$ and $\omega$ are time-independent, and derivatives are with respect to $\tau$. Since equation \eqref{EquationOfMotionRedefinedTimeCo} is a differential equation $E(\tau,q,\dot{q},\ddot{q})$, it may be substituted into equation \eqref{SecondHelmholtzCondition}, to ensure that the Helmholtz conditions are satisfied:
\begin{align}
    \left( \frac{d^2 \tau}{dt^2} \right) + 2 \alpha \frac{d\tau}{dt} = \frac{d}{d\tau} \left[ \left( \frac{d\tau}{dt} \right)^2 \right].
    \label{ConstraintEquationForTau}
\end{align}
This condition is trivially satisfied for $\alpha = 0$ (the undamped oscillator), and we consider $\alpha \not = 0$ for the remainder of this work. In that case, equation \eqref{ConstraintEquationForTau} is generally satisfied by
\begin{align}
    \tau = K e^{2 \alpha t}
    - \tau_0, \label{TauDefinition} 
\end{align}
where $K$ and $\tau_0$ are constants of integration. $K$ must be non-zero for the transform to be defined, but $\tau_0$ corresponds only to a time translation of our system $\tau \rightarrow \tau + \tau_0$. Since this has no physical implications, we set $\tau_0 = 0$ for the remainder of this work. \\\\
Observe that equation \eqref{ConstraintEquationForTau} is independent of $g_i(q,t)$, so it is possible to write a Lagrangian for any linearly damped system. However, we start by considering the undriven damped harmonic oscillator, for which the equation of motion in $\tau$ is 
\begin{align}
    4 \alpha^2 \tau^2 \frac{d^2 q}{d\tau^2} + 8 \alpha^2 \tau \frac{dq}{d\tau} + \omega^2 q = 0.
    \label{DampedOscillatorInTau}
\end{align}
It is straightforward to show that \eqref{DampedOscillatorInTau} is the Euler-Lagrange equation corresponding to the Lagrangian
\begin{align}
    \mathcal{L} = \frac{1}{2} \left[ 4 \alpha^2 \tau^2 \dot{q}^2 - \omega^2 q^2 \right]. 
    \label{LagrangianinTau}
\end{align}
The canonical momentum $p$ for $\mathcal{L}$ is 
\begin{align}
    p = \frac{\partial L}{\partial \dot{q}} = 4 \alpha^2 \tau^2 \dot{q}, 
    \label{CanonicalMomentuminTau}
\end{align}
and thus the classical Hamiltonian $H$ for our system is
\begin{align}
    H = \frac{1}{2} \left[ \frac{1}{4 \alpha^2 \tau^2} p^2 + \omega^2 q^2 \right].
    \label{HamiltonianinTau}
\end{align}
It follows from equation \eqref{TauDefinition} that $\tau > 0$ for $t > -\infty$, so $H$ is defined for all time and $\alpha > 0$. So the resulting classical Hamiltonian, equation \eqref{HamiltonianinTau} produces equation \eqref{DampedHarmonicOscillatorEquation} as its equation of motion. To the best of our knowledge, equation \eqref{HamiltonianinTau} has not been found in the earlier literature on the quantization of the damped harmonic oscillator. Nor has this procedure for deriving a Hamiltonian for a dissipative system been proposed in the earlier literature on the quantization of dissipative systems. 
\section{Canonical Quantization}
Canonical quantization is now carried out in the usual way by promoting $q$ and $p$ to their corresonding self-adjoint operators satisfying $\left[ \hat{q}, \hat{p} \right] = i\hbar$. We now define, in the Schr\"{o}dinger picture, the annihilation operator
\begin{align}
    \hat{a}(\tau) = \frac{1}{\sqrt{2\hbar}} \left[ \sqrt{2 \omega \alpha \tau} \hat{q} + i \frac{1}{\sqrt{2 \omega \alpha \tau}} \hat{p} \right],
\end{align}
which satisfies $\left[ \hat{a}(\tau), \hat{a}^{\dagger}(\tau) \right] = 1$ for all $\tau$, where $\hat{a}^{\dagger}(\tau)$ is the corresponding creation operator. Our quantized Hamiltonian may thus be written
\begin{align}
    \hat{H}(\tau) = \frac{\hbar \omega}{2 \alpha \tau} \left[ \hat{a}^{\dagger}(\tau) \hat{a}(\tau) + \frac{1}{2} \right].
    \label{HamiltonianInCreationAndAnnihilation}
\end{align}
Using the ground state vector $| 0, \tau \rangle$ which satisfies $\hat{a}(\tau) | 0, \tau \rangle = 0$, the Fock basis vectors are constructed as $| n, \tau \rangle = ( 1/\sqrt{n!} ) \left( \hat{a}^{\dagger}(\tau)\right)^n | 0, \tau \rangle$ for positive integer $n$. The Fock basis vectors are eigenstates of the Hamiltonian $\hat{H}(\tau) | n, \tau \rangle = E_n(\tau) | n, \tau \rangle$, where
\begin{align}
    E_n \left( \tau \right) = \frac{\hbar \omega}{2 \alpha \tau} \left( n + \frac{1}{2} \right).
\end{align}
Thus, in the original time coordinate $t$, the energy spectrum of our damped harmonic oscillator is
\begin{align}
    E_n \left( t \right) = \frac{\hbar \omega}{2 \alpha K} e^{-2 \alpha t} \left( n + \frac{1}{2} \right).
\end{align}
Consistent with equation \eqref{DeguchiEnergySpectrum}, the energy eigenvalues constitute an equally spaced ladder, and decrease exponentially with time. Note that a choice of $K = 1/2 \alpha$ ensures that $E_n$ reproduces exactly the harmonic oscillator spectrum at $t = 0$. The eigenvalues decrease as $\exp \left( - 2 \alpha t \right)$, which is the same rate as that expected for a corresponding classical oscillator.

\section{Position-space energy eigenfunctions}
\begin{figure}[t!]

\subfloat[]{%
  \includegraphics[clip,width=\columnwidth]{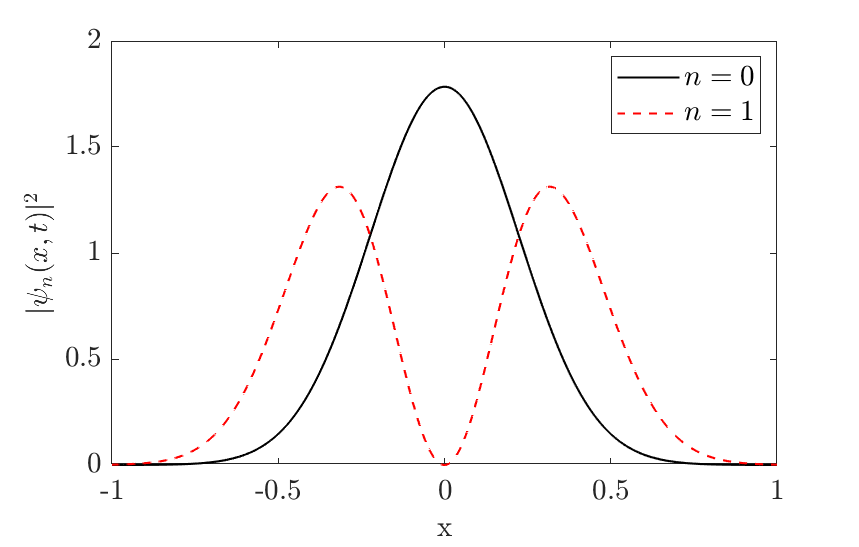} \label{fig:Fig1a}
}

\subfloat[]{%
  \includegraphics[clip,width=\columnwidth]{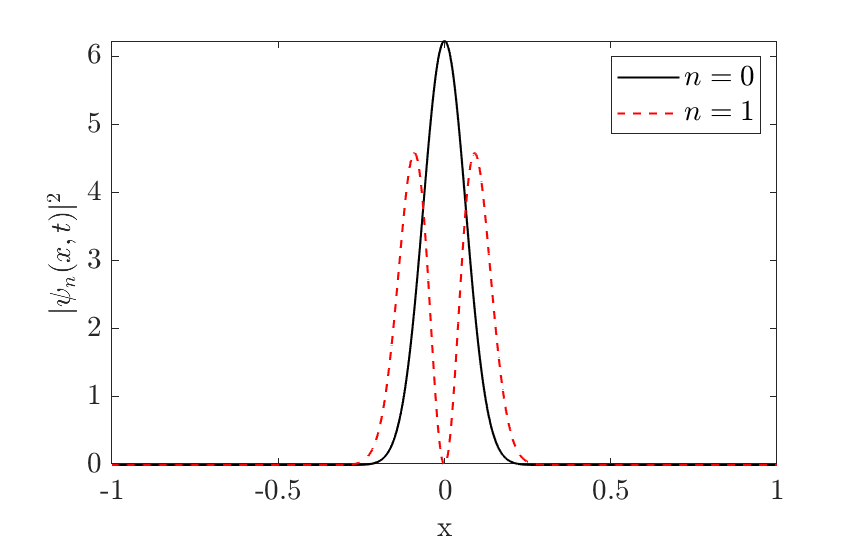} \label{fig:Fig1b}
}

\caption{Figures \ref{fig:Fig1a} and \ref{fig:Fig1b} show $|\psi_n(x,t)|^2$ for equation \eqref{nthwavefunctionUS} for $n=0,1$ at $t=0$ and $t = 250$ respectively. }
\label{fig:Fig1Total}
\end{figure}

We define the Fock state $\ket{n, \tau}$ in the position basis $ \ket{x}$
\begin{align}
    | n, \tau \rangle = \int dx \  \psi_n(x,\tau) \ket{x},
\end{align}
where the $\psi_n(x,\tau)$ are the position-space wavefunctions. Having defined the annihiliation operator $\hat{a}(\tau)$, following a standard approach Ref. \cite{griffiths_introduction_2005} the eigenfunctions
\begin{align}
\begin{split}
    \psi_n (x,\tau) = & \left( \frac{2 \omega \alpha \tau}{\pi \hbar} \right)^{1/4} \frac{1}{\sqrt{2^n n!}} \\ 
    & \times H_n \left( \sqrt{\frac{2 \omega \alpha \tau}{\hbar}} x \right) \exp \left( - \frac{\omega \alpha \tau}{ \hbar} x^2 \right),
\end{split}
\end{align}
are obtained, where $H_n$ is the $n$th Hermite polynomial. Writing these wavefunctions in the original time coordinate then gives
\begin{align}
\begin{split}
    \psi_n(x,t) = &\left( \frac{2 \omega \alpha K}{\pi \hbar} \right)^{1/4} \frac{1}{\sqrt{2^n n!}} \\
    & \times H_n \left( \sqrt{\frac{2 \omega \alpha K}{\hbar}} \exp \left( \alpha t \right) x \right)  \\
    & \times \exp \left( \frac{\alpha t}{2} - \frac{\omega \alpha K}{ \hbar} x^2 \exp \left( 2 \alpha t \right) \right).
\end{split}
\label{nthwavefunctionUS}
\end{align}
Observe that, up to a phase factor, these are the same wavefunctions as found in Ref. \cite{deguchi_quantization_2020} for $K = m/2\alpha$ (where $m$ is the mass of the oscillator). In Figure \ref{fig:Fig1a}, we plot $|\psi_n(x,t)|^2$ for equation \eqref{nthwavefunctionUS} at $t=0$ for the $n = 0$ and $n=1$ cases, which correspond to the initial position-space wavefunctions of the ground state and first excited state, for $\omega = 1$, $\alpha = 0.005$, $\hbar = 1$ and $K = 1/2\alpha$. \\\\
In Figure \ref{fig:Fig1b} we plot the $n=0$ and $n=1$ wavefunctions at $t=250$ for equation \eqref{nthwavefunctionUS} with the same choice of constants. As $t$ increases, observe that $|\psi_n(x,t)|^2$ becomes increasingly localised around the origin $x = 0$, as the oscillator's motion is damped.

\section{Solutions to the Schr\"{o}dinger Equation}
We now calculate the evolution of a state obeying the time-dependent Hamiltonian equation \eqref{HamiltonianInCreationAndAnnihilation} in the Schr\"{o}dinger picture. In particular, our time-dependent wavefunctions obey the eigenvalue equation
\begin{align}
    \hat{H}(\tau) \psi_n(x,\tau) = E_n(\tau) \psi_n(x,\tau).
\end{align}
As the energy levels of our Hamiltonian are nondegenerate, the general solutions $\Psi(x,\tau)$ for the time-dependent Schr\"{o}dinger equation is found in the standard way to give, in terms of $\tau$,   
\begin{align}
    \Psi(x,\tau) = \sum_n c_n(\tau) \psi_n(x,\tau) e^{i\theta_n (\tau)},
    \label{SchrodingerSolutionTau}
\end{align}
where $|c_n(\tau)|^2$ is the probability of being in the $n$th energy level at time $\tau$, and 
\begin{align}
    \theta_n \left( \tau \right) = - \frac{1}{\hbar} \int_K^{\tau} E_n(\tau') d\tau'.
    \label{ThetaDefinition}
\end{align}
Here, the $c_m$ obey
\begin{align}
\begin{split}
    \frac{dc_m}{d\tau} = &- c_m \bra{m, \tau} \frac{d}{d\tau} \ket{m, \tau} \\
    & - \sum_{n\not=m} c_n \frac{\bra{m, \tau} \frac{dH}{d\tau} \ket{n, \tau}}{E_n - E_m} e^{i \left( \theta_n - \theta_m \right)},
    \end{split}
    \label{EvolutionOfCM}
\end{align}
where $\ket{m, \tau}$ are the Fock basis states defined before. The integral in equation \eqref{ThetaDefinition} is integrated from $K$, corresponding to $\tau (t =0 )$. Then, for the Hamiltonian given by equation \eqref{HamiltonianinTau}, equation \eqref{EvolutionOfCM} becomes
\begin{align}
    \begin{split}
        \frac{dc_m}{d\tau} = & \frac{1}{4\tau} \left[ c_{m-2} \sqrt{m(m-1)} \left( \frac{\tau}{K} \right)^{i\omega / \alpha} \right. \\
        & \left. - c_{m+2} \sqrt{(m+2)(m+1)} \left( \frac{\tau}{K} \right)^{-i\omega / \alpha} \right].
    \end{split}
 \label{CMEquationInTau}
\end{align}
Transforming back to the original time coordinate $t$ gives
\begin{align}
    \begin{split}
        \frac{dc_m}{dt} = &\frac{\alpha}{2} \left[ c_{m-2} \sqrt{m(m-1)} e^{i2\omega t} \right. \\
        & \left. - c_{m+2} \sqrt{(m+2)(m+1)} e^{-i2\omega t} \right],
    \end{split}
\label{CMEquationInT}
\end{align}
for the time rate of change of the coefficients. Equation \eqref{CMEquationInT} may be solved as described in Ref. \cite{deguchi_quantization_2020}, and the references therein, for initial condition $c_m(0) = \delta_{m,n}$, where $\ket{n, 0}$ is the initial state of our system. This is achieved by solving
\begin{align}
    \frac{\partial G}{\partial t} = - \left[ \frac{\alpha}{2} \left( \frac{\partial^2}{\partial q^2} - q^2 \right) + i \omega q \frac{\partial}{\partial q} \right] G,
\end{align}
for $G(q,t) \triangleq \sum_j q^j e^{-ij \omega t / 2} c_j(t) / \sqrt{j!}$ with initial conditions $G(q,0) = q^n /\sqrt{n!}$ and $G(0,t) = c_0(t)$. The $c_m(\tau)$ can then be used to find the probability of the system being found in any eigenstate, after initially being in the $n$th energy level.  We consider here the cases $n = 0$ and $n=2$ in particular. Observe that interestingly equation \eqref{CMEquationInT} only couples modes of the same parity, so it is sufficient to consider only the even-order modes. 
\subsection{Case $n = 0$}
\begin{figure}[t!]

\subfloat[]{%
  \includegraphics[clip,width=\columnwidth]{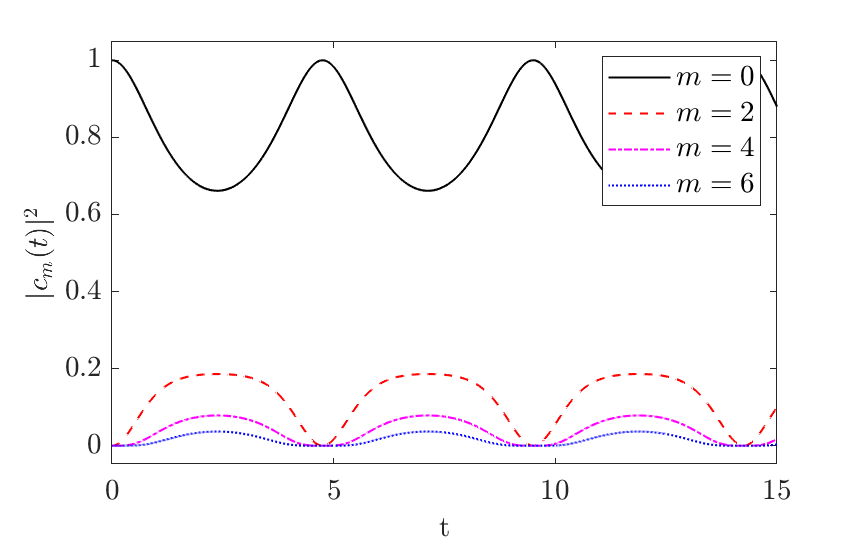} \label{fig:Fig2a}
}

\subfloat[]{%
  \includegraphics[clip,width=\columnwidth]{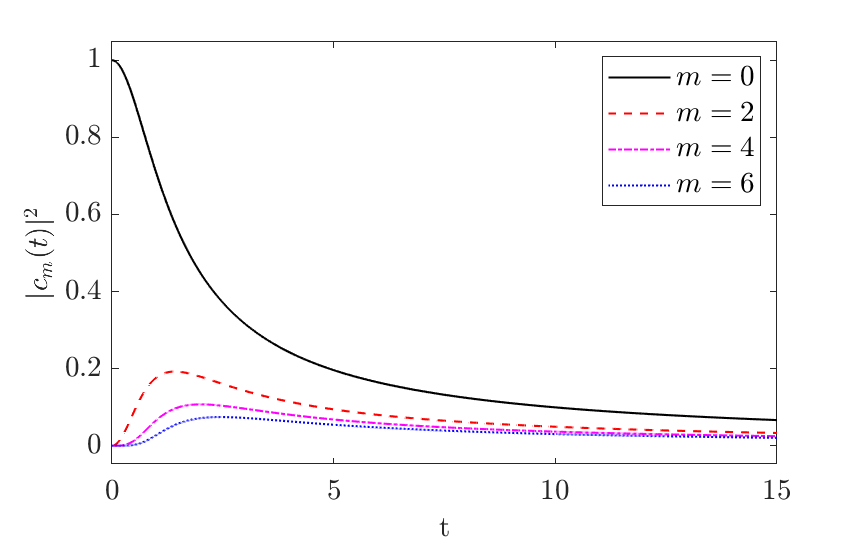} \label{fig:Fig2b}
}

\subfloat[]{%
  \includegraphics[clip,width=\columnwidth]{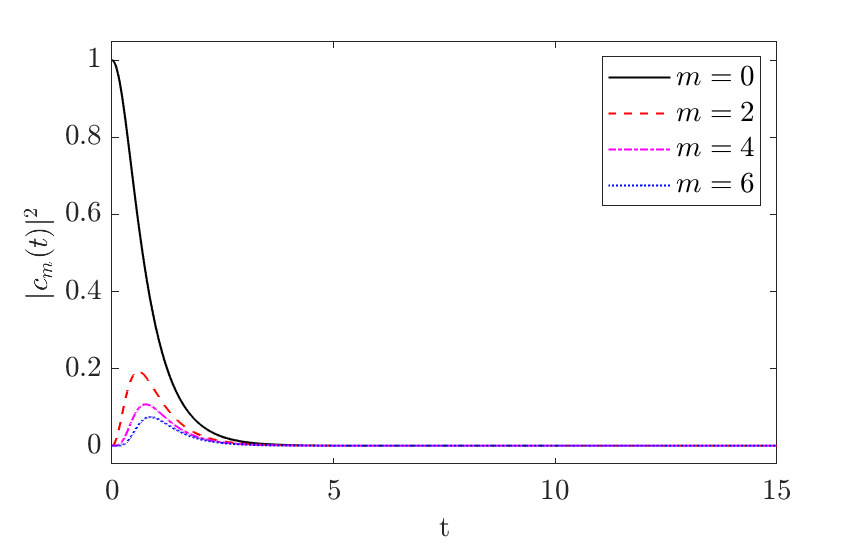} \label{fig:Fig2c}
}

\caption{For an initial state of $\ket{0, 0}$, we plot $|c_m(t)|^2$ for $m = 0, 2, 4, 6$ in the underdamped case (Figure \ref{fig:Fig2a}), the critically damped case (Figure \ref{fig:Fig2b}), and the overdamped case (Figure \ref{fig:Fig2c}). }
\label{fig:Fig2Total}
\end{figure}

In the case $n=0$, where the system in initially in the ground state, $c_m(t)$ may be written
\begin{align}
    c_m(t) = \begin{cases}
    & \frac{(m-1)!!}{\sqrt{m!}} \sqrt{\xi} e^{i(m+1/2) \omega t} \\
    & \times \frac{\left( \sinh \left( \xi \alpha t \right) \right)^{m/2}}{\left( \cosh \left( \zeta + \xi \alpha t \right) \right)^{(m+1)/2}}, \\
    & \text{for } m = 2k, \\ 
    & \\
    & 0, \text{ for } m = 2k + 1,
    \end{cases}
\end{align}
where $\sum_m |c_m(t)|^2 = 1$, $k \in \mathbb{Z}$, and
\begin{subequations}
\begin{align}
    \xi & = \sqrt{1 - \omega^2/\alpha^2}, \label{XiDefinition}\\
    e^{\pm \zeta} &= \xi \pm i\omega/\alpha.
\end{align}
\label{XiandZetaDefinitions}
\end{subequations}
Critical damping occurs when $\xi = 0$ which, from equation \eqref{XiDefinition}, gives the condition for critical damping as $\alpha = \omega$, consistent with the result for an equivalent classical system. Note that this condition differs from the result of Ref. \cite{deguchi_quantization_2020} ($\alpha = \omega (\sqrt{5}-1)/2$).\\\\
In Figure \ref{fig:Fig2Total} we show $|c_m(t)|^2$ with $m = 0,2,4,6$ for (a) the underdamped case ($\alpha< \omega$), (b) the case of critical damping ($\alpha= \omega$), and (c) the overdamped case ($\alpha> \omega$) with $\omega = 1$ and $\alpha$ defined accordingly.\\\\
In case a), $\xi$ is imaginary and $c_m(t)$ is oscillatory. In Figure \ref{fig:Fig2a}, we show the underdamped case, with $\alpha = 0.75 <  \omega$. The transition probabilities oscillate with period $\pi/ |\alpha \xi| = \pi/\sqrt{\omega^2-\alpha^2}$. \\\\
In case b), $\xi = 0$. Expanding about $\xi = 0$, we then obtain
\begin{align}
\begin{split}
    c_m(t) = \frac{(m-1)!!}{\sqrt{m!}} e^{i(m+1/2)\omega t} \frac{\left( \omega t \right)^{n/2}}{\left( 1 + i \omega t \right)^{\left( n+1 \right)/2}}
    \end{split}.
\end{align}
In Figure \ref{fig:Fig2b}, we show this critically damped case, with $\alpha = \omega = 1$. The transition probability $|c_m(t)|^2$ decreases monotonically for the ground state $m=0$, whilst for $m \geq 2$ it increases to a local maximum before also decreasing monotonically.\\\\
In Figure \ref{fig:Fig2c} we show the overdamped case $\alpha = 2 > \omega$. Now $\xi \not =0$ and has only real components, so we observe similar behaviour to the case of critical damping except for the expected faster rate of decay.
\subsection{Case $n = 2$}
When the initial state is the second excited state of the system, $n = 2$, the transition amplitudes are
\begin{align}
    c_m(t) = \begin{cases}
    & \frac{(m-1)!!}{\sqrt{2m!}} \sqrt{\xi} e^{i(m+1/2) \omega t} \\
    & \times \frac{\left( \sinh \left( \xi \alpha t  \right) \right)^{m/2}}{\left( \cosh \left( \zeta + \xi \alpha t  \right) \right)^{(m+3)/2}} \\
    & \times \left( \frac{m \xi^2}{\sinh \left( \xi \alpha t  \right)} - \sinh \left( \xi \alpha t  \right) \right), \\
    & \text{for } m = 2k, \\
    & \\
    & 0, \text{ for } m = 2k + 1,
    \end{cases}
\end{align}
which again satisfies the normalization condition $\sum_m |c_m(t)|^2 = 1$, with the parameters $\xi$ and $\zeta$ defined as before.\\\\
In Figure \begin{figure}[t!]

\subfloat[]{%
  \includegraphics[clip,width=\columnwidth]{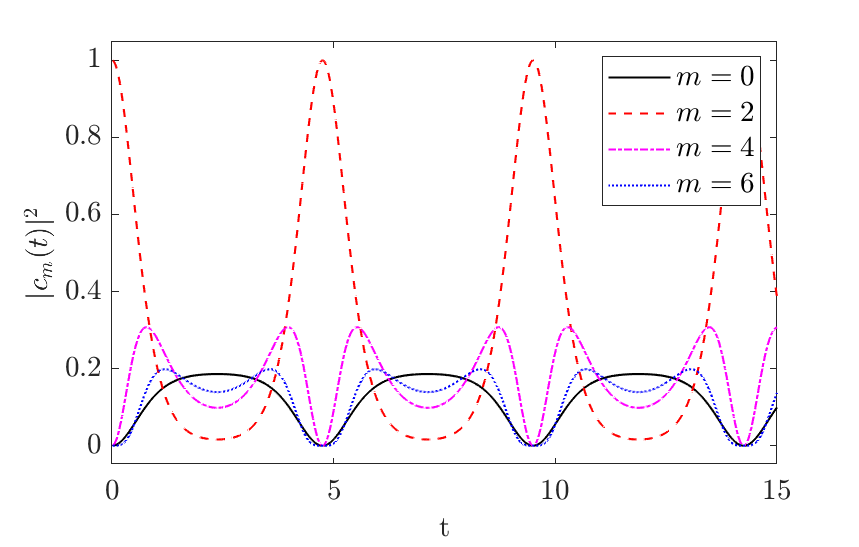} \label{fig:Fig3a}
}

\subfloat[]{%
  \includegraphics[clip,width=\columnwidth]{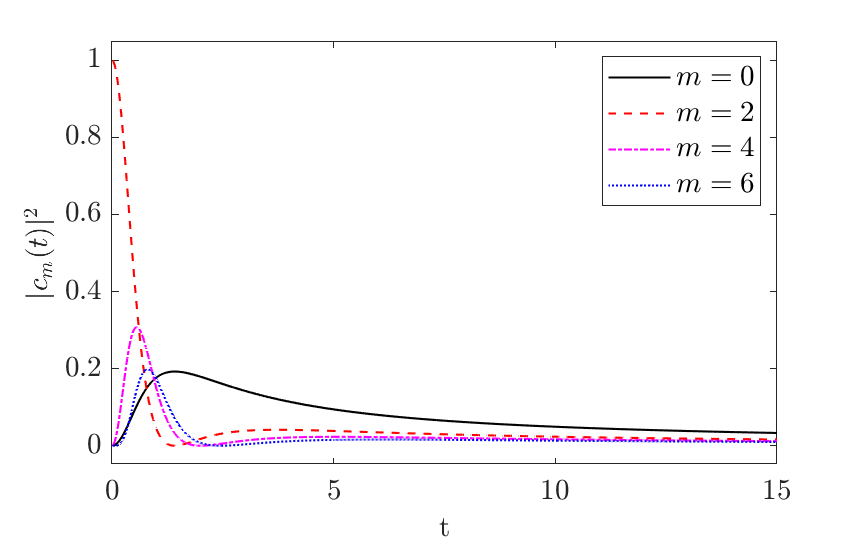} \label{fig:Fig3b}
}

\subfloat[]{%
  \includegraphics[clip,width=\columnwidth]{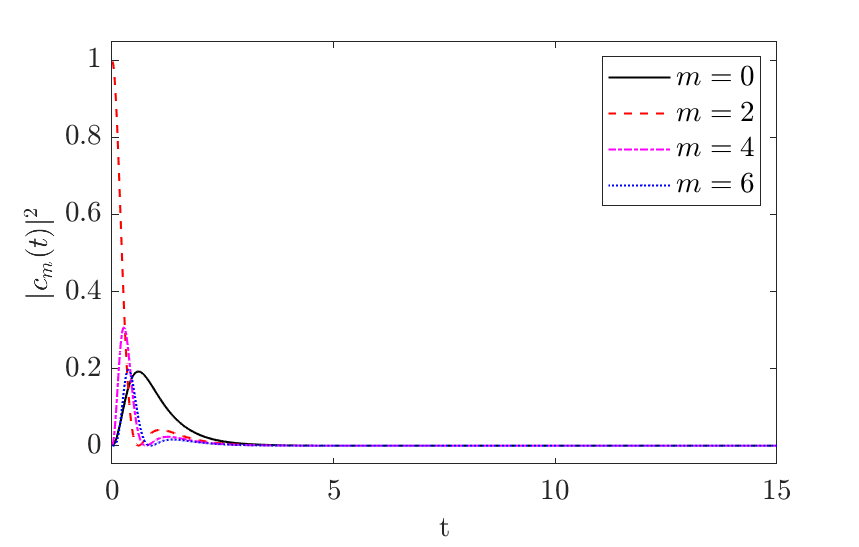} \label{fig:Fig3c}
}

\caption{For an initial state of $\ket{2, 0}$, we plot $|c_m(t)|^2$ for $m = 0, 2, 4, 6$ in the underdamped case (Figure \ref{fig:Fig3a}), the critically damped case (Figure \ref{fig:Fig3b}), and the overdamped case (Figure \ref{fig:Fig3c}). }
\label{fig:Fig3Total}
\end{figure} \ref{fig:Fig3Total}, we plot $|c_m(t)|^2$ with $m=0,2,4,6$ for (a) the underdamped case ($\alpha < \omega$), (b) the case of critical damping ($\alpha = \omega$), and (c) the overdamped case ($\alpha > \omega$) with $\omega = 1$ and $\alpha$ defined accordingly. \\\\
Figure \ref{fig:Fig3a} shows $|c_m(t)|^2$ for $\alpha = 0.75 < \omega$. As before, we observe oscillations in the transition probabilities with period $\pi/|\alpha \xi| = \pi/\sqrt{\omega^2 - \alpha^2}$. However, we observe ``beating'' between $\alpha$ and $\omega$ in the modes $m > n$. This beating is not present in the $n=0$ case considered earlier.\\\\
As before, in the critically damped case $\xi = 0$, so following Ref. \cite{deguchi_quantization_2020} we expand around $\xi = 0$ to obtain
\begin{align}
\begin{split}
    c_m(t) = & \frac{\left( m-1\right)!!}{\sqrt{2m!}} e^{i(m+1/2)\omega t} \left( \frac{m}{\omega t} - \omega t \right) \\
    & \times \frac{\left( \omega t \right)^{m/2}}{\left( 1 + i \omega t \right)^{\left( m + 3 \right)/2}}.
\end{split}
\end{align}
In Figure \ref{fig:Fig3b} we show the critically damped case with $\alpha = \omega = 1$. In this case it is the initially occupied $m=2$ mode which decreases monotonically, whilst the $m=0$ mode increases before decreasing. \\\\
In Figure \ref{fig:Fig3c}, we show the overdamped case $\alpha = 2 > \omega$. As before, we observe qualitatively similar behaviour to the critical damping case, except for the expected greater rate of decay.
\section{Examples of Driven Damped Harmonic Oscillators}
As foreshadowed in the introduction, this quantization of the damped harmonic oscillator has many potential applications. In what follows the quantization procedure is applied to two of the simplest superconducting qubits, the phase qubit and flux qubit, and provides an example of the application of the quantization procedure in the case of a driven damped harmonic oscillator.  
The coordinate of interest here is $\delta$, the phase of the current through the Josephson junction, and as is well known, the equations of motion for these qubits in terms of this dynamical variable are determined through the use of the RCSJ model and Kirchhoff's circuital laws and may be written
\begin{subequations}
\begin{align}
    \frac{\hbar C}{2e} \frac{\partial^2 \delta}{\partial t^2} + \frac{\hbar}{2eR} \frac{\partial \delta}{\partial t} + \left( I_0 \sin \delta - I \right) & = 0, \label{EqPhaseQubit} \\
    \frac{\hbar C}{2e} \frac{\partial^2 \delta}{\partial t^2} + \frac{\hbar}{2eR} \frac{\partial \delta}{\partial t} + I_0 \sin \delta + \frac{\hbar}{2eL} \left( \delta - \delta_X \right) & = 0. \label{EqFluxQubit}%
\end{align}
\label{EqQubitsTotal}%
\end{subequations}
Here, $C$ and $R$ are the capacitance and resistance of the circuit respectively, and $I_0$ is the critical current of the Josephson junction. $I$ is the current bias in the phase qubit, and $L$ and $\delta_X$ respectively the inductance and a dimensionless parameter for the flux through the flux qubit. Neither equation \eqref{EqPhaseQubit} nor \eqref{EqFluxQubit} satisfy the second Helmholtz condition (equation \eqref{SecondHelmholtzCondition}). Thus, in both cases, we introduce the dimensionless time coordinate
\begin{align}
    \tau = K \exp \left( \frac{t}{CR} \right),
\end{align}
where $K$ is a constant of integration \footnote{As justified earlier, here we set $\tau_0 = 0$.}, obtaining
\begin{subequations}
\begin{align}
    \begin{split}
        \left( \frac{\tau}{CR} \right)^2 \frac{d^2 \delta}{d\tau^2} &+ \frac{2 \tau}{ C^2 R^2} \frac{d \delta}{d\tau} \\
        &+ \frac{2e}{\hbar C} \left( I_0 \sin \delta - I \right)  = 0, 
    \end{split}\\
    \begin{split}
        \left( \frac{\tau}{CR} \right)^2 \frac{d^2 \delta}{d\tau^2} & + \frac{2\tau}{C^2 R^2} \frac{d \delta}{d\tau} \\
        & + \frac{2e}{\hbar C} I_0 \sin \delta + \frac{1}{LC}\left( \delta - \delta_X \right)  = 0,
    \end{split}
\end{align}
\end{subequations}
for the phase and flux qubits respectively. We can define the same canonical momentum $\pi$ for each case
\begin{align}
    \pi = \left( \frac{\tau}{CR} \right)^2 \dot{\delta},
\end{align}
and obtain Hamiltonians for each system. These are $H_P$ and $H_F$
\begin{subequations}
\begin{align}
    H_P = & \frac{1}{2} \left[ \left( \frac{CR}{\tau} \right)^2 \pi^2 - \frac{4e}{\hbar C} \left( I_0 \cos \delta + I \delta \right) \right], \\
    \begin{split}
    H_F = & \frac{1}{2} \left[ \left( \frac{CR}{\tau} \right)^2 \pi^2  - \frac{4e}{\hbar C} I_0 \cos \delta \right. \\
    & \left. + \frac{1}{LC} \left( \delta^2 - 2 \delta_X \delta \right) \right],
    \end{split}
\end{align}
\end{subequations}
for the phase and flux qubit respectively. Quantization is now achieved through the standard method of promoting $\delta$ and $\pi$ to their corresponding self-adjoint operators satisfying the commutator $[ \hat{\delta}, \hat{\pi} ] = i\hbar$. In the regime where $\hat{\delta}$ is small, we use the series expansion
\begin{align}
    \cos \hat{\delta} \approx 1 - \frac{1}{2} \hat{\delta}^2,
\end{align} 
to write
\begin{subequations}
\begin{align}
    \hat{H}_P = &\frac{1}{2} \left[ \frac{1}{\alpha_P^2 \tau^2} \hat{\pi}^2 + \Omega_P^2 \left( \hat{\delta}^2 - 2 \frac{I}{I_0} \hat{\delta} - 2 \right) \right],\\
    \begin{split}
        \hat{H}_F = & \frac{1}{2} \left[ \frac{1}{\alpha_F^2 \tau^2} \hat{\pi}^2 + \Omega_F^2 \right. \\
        & \left. \times \left( \hat{\delta}^2 - \frac{2\delta_X}{LC \Omega_F^2} \hat{\delta} - 2 \frac{\Omega_P^2}{\Omega_F^2} \right) \right],
    \end{split}
\end{align}
\end{subequations}
where we have defined
\begin{subequations}
\begin{align}
    \alpha_P = & \alpha_F = \frac{1}{CR}, \\
    \Omega_P = & \sqrt{\frac{2eI_0}{\hbar C}}, \\
    \Omega_F = & \sqrt{\Omega_P^2 + \frac{1}{LC}}.
\end{align}
\end{subequations}
The translated operators
\begin{subequations}
\begin{align}
    \hat{\delta}_P &= \hat{\delta} - \frac{I}{I_0}, \\
    \hat{\delta}_F &= \hat{\delta} - \frac{\delta_X}{LC \Omega_F^2}, 
\end{align}
\end{subequations}
obey the commutation relations $[ \hat{\delta}_P, \hat{\pi} ] = [ \hat{\delta}_F, \hat{\pi} ] = i\hbar$. In terms of these new canonically conjugate pairs, we obtain the Hamiltonians
\begin{subequations}
\begin{align}
    \hat{H}_P = &\frac{1}{2} \left[ \frac{1}{\alpha_P^2 \tau^2} \hat{\pi}^2 + \Omega_P^2 \hat{\delta}_P^2 \right] - \left( \Omega_P^2 + \left( \frac{I}{I_0} \right)^2 \right),\\
    \begin{split}
        \hat{H}_F = &\frac{1}{2} \left[ \frac{1}{\alpha_P^2 \tau^2} \hat{\pi}^2 + \Omega_F^2 \hat{\delta}_F^2 \right] \\
        & - \left( \Omega_P^2 + \left( \frac{\delta_X}{LC \Omega_F^2} \right)^2 \right).
    \end{split}
\end{align}
\end{subequations}
From our earlier results, the energy spectrums $E_n^{(P)}(t)$ and $E_n^{(F)}(t)$ for each of the qubits are
\begin{subequations}
\begin{align}
\begin{split}
    E_n^{(P)}(t) = \frac{\hbar \Omega_P}{\alpha_P K} e^{- \alpha_P t} & \left( n + \frac{1}{2} \right) \\
    & - \left( \Omega_P^2 + \left( \frac{I}{I_0} \right)^2 \right), 
\end{split} \\
\begin{split}
    E_n^{(F)}(t) = \frac{\hbar \Omega_F}{\alpha_F K} e^{- \alpha_F t} & \left( n + \frac{1}{2} \right) \\
    &- \left( \Omega_P^2 + \left( \frac{\delta_X}{LC \Omega_F^2} \right)^2 \right).
\end{split}
\end{align}
\end{subequations}
The associated conditions for critical damping are 
\begin{subequations}
\begin{align}
    R &= \sqrt{\frac{\hbar}{8 e I_0 C}}, \\
    R &= \sqrt{\frac{\hbar L}{4C \left( 2 e I_0 L + \hbar \right)}},
\end{align}
\end{subequations}
for the phase and flux qubits respectively. 
\section{Conclusions}
We have proposed a new approach to the quantization of driven and undriven linearly damped harmonic oscillators via a change of time coordinate to produce classical equations of motion that satisfy the Helmholtz conditions. The resulting quantum model is qualitatively similar to previous approaches, such as Ref. \cite{deguchi_quantization_2020}, since it predicts an equally spaced ladder of energy eigenvalues decaying at rate $\exp(-2\alpha t)$. However, the result of our quantization predicts a critical damping parameter $\alpha = \omega$, which is the same as for an equivalent classical oscillator. Thus, our quantized system satisfies the correspondence principle. \\\\
It has been shown that a system which is initially in a single eigenstate transitions between eigenstates of the same parity. Additionally, for systems initially in the second excited state the quantization predicts different dynamics and the presence of beating when $\alpha < \omega$, vanishing at the point of critical damping.
The energy eigenvalues and the critical damping points have been found for the cases of the simplest phase and flux qubits.\\\\
Importantly, this new quantization method for the damped harmonic oscillator requires only a single classical dynamical variable and is therefore easily generalised. In particular, this approach should also be applicable to systems with time-dependent damping $\alpha(t)$, in which case the constraint equation for $\tau$ becomes
\begin{align}
    \left( \frac{d^2 \tau}{dt^2} \right) + 2 \alpha \left( t \right) \frac{d\tau}{dt} = \frac{d}{d\tau} \left[ \left( \frac{d\tau}{dt} \right)^2 \right],
\end{align}
which will admit different solutions depending upon the functional form of $\alpha$.

\section{Acknowledgements}
The authors would like to thank Murray Batchelor for fruitful discussions in the preparation of this manuscript. The authors would also like to thank Tin Sulejmanpasic for pointing out an error in the original manuscript.

\bibliographystyle{bibsty}
\bibliography{bib.bib}

\begin{thebibliography}{37}%
\makeatletter
\providecommand \@ifxundefined [1]{%
 \@ifx{#1\undefined}
}%
\providecommand \@ifnum [1]{%
 \ifnum #1\expandafter \@firstoftwo
 \else \expandafter \@secondoftwo
 \fi
}%
\providecommand \@ifx [1]{%
 \ifx #1\expandafter \@firstoftwo
 \else \expandafter \@secondoftwo
 \fi
}%
\providecommand \natexlab [1]{#1}%
\providecommand \enquote  [1]{``#1''}%
\providecommand \bibnamefont  [1]{#1}%
\providecommand \bibfnamefont [1]{#1}%
\providecommand \citenamefont [1]{#1}%
\providecommand \href@noop [0]{\@secondoftwo}%
\providecommand \href [0]{\begingroup \@sanitize@url \@href}%
\providecommand \@href[1]{\@@startlink{#1}\@@href}%
\providecommand \@@href[1]{\endgroup#1\@@endlink}%
\providecommand \@sanitize@url [0]{\catcode `\\12\catcode `\$12\catcode
  `\&12\catcode `\#12\catcode `\^12\catcode `\_12\catcode `\%12\relax}%
\providecommand \@@startlink[1]{}%
\providecommand \@@endlink[0]{}%
\providecommand \url  [0]{\begingroup\@sanitize@url \@url }%
\providecommand \@url [1]{\endgroup\@href {#1}{\urlprefix }}%
\providecommand \urlprefix  [0]{URL }%
\providecommand \Eprint [0]{\href }%
\providecommand \doibase [0]{http://dx.doi.org/}%
\providecommand \selectlanguage [0]{\@gobble}%
\providecommand \bibinfo  [0]{\@secondoftwo}%
\providecommand \bibfield  [0]{\@secondoftwo}%
\providecommand \translation [1]{[#1]}%
\providecommand \BibitemOpen [0]{}%
\providecommand \bibitemStop [0]{}%
\providecommand \bibitemNoStop [0]{.\EOS\space}%
\providecommand \EOS [0]{\spacefactor3000\relax}%
\providecommand \BibitemShut  [1]{\csname bibitem#1\endcsname}%
\let\auto@bib@innerbib\@empty
\bibitem [{\citenamefont {Kjaergaard}\ \emph {et~al.}(2020)\citenamefont
  {Kjaergaard}, \citenamefont {Schwartz}, \citenamefont {Braumüller},
  \citenamefont {Krantz}, \citenamefont {Wang}, \citenamefont {Gustavsson},\
  and\ \citenamefont {Oliver}}]{kjaergaard_superconducting_2020}%
  \BibitemOpen
  \bibfield  {author} {\bibinfo {author} {\bibfnamefont {M.}~\bibnamefont
  {Kjaergaard}}, \bibinfo {author} {\bibfnamefont {M.~E.}\ \bibnamefont
  {Schwartz}}, \bibinfo {author} {\bibfnamefont {J.}~\bibnamefont
  {Braumüller}}, \bibinfo {author} {\bibfnamefont {P.}~\bibnamefont {Krantz}},
  \bibinfo {author} {\bibfnamefont {J.~I.-J.}\ \bibnamefont {Wang}}, \bibinfo
  {author} {\bibfnamefont {S.}~\bibnamefont {Gustavsson}}, \ and\ \bibinfo
  {author} {\bibfnamefont {W.~D.}\ \bibnamefont {Oliver}},\ }\href {\doibase
  10.1146/annurev-conmatphys-031119-050605} {\bibfield  {journal} {\bibinfo
  {journal} {Annual Review of Condensed Matter Physics}\ }\textbf {\bibinfo
  {volume} {11}},\ \bibinfo {pages} {369} (\bibinfo {year} {2020})},\ \bibinfo
  {note} {\_eprint:
  https://doi.org/10.1146/annurev-conmatphys-031119-050605}\BibitemShut
  {NoStop}%
\bibitem [{\citenamefont {Devoret}\ and\ \citenamefont
  {Schoelkopf}(2013)}]{devoret_superconducting_2013}%
  \BibitemOpen
  \bibfield  {author} {\bibinfo {author} {\bibfnamefont {M.~H.}\ \bibnamefont
  {Devoret}}\ and\ \bibinfo {author} {\bibfnamefont {R.~J.}\ \bibnamefont
  {Schoelkopf}},\ }\href {\doibase 10.1126/science.1231930} {\bibfield
  {journal} {\bibinfo  {journal} {Science}\ }\textbf {\bibinfo {volume}
  {339}},\ \bibinfo {pages} {1169} (\bibinfo {year} {2013})},\ \bibinfo {note}
  {publisher: American Association for the Advancement of Science Section:
  Review}\BibitemShut {NoStop}%
\bibitem [{\citenamefont {Mitra}\ \emph {et~al.}(2008)\citenamefont {Mitra},
  \citenamefont {Strauch}, \citenamefont {Lobb}, \citenamefont {Anderson},
  \citenamefont {Wellstood},\ and\ \citenamefont
  {Tiesinga}}]{mitra_quantum_2008}%
  \BibitemOpen
  \bibfield  {author} {\bibinfo {author} {\bibfnamefont {K.}~\bibnamefont
  {Mitra}}, \bibinfo {author} {\bibfnamefont {F.~W.}\ \bibnamefont {Strauch}},
  \bibinfo {author} {\bibfnamefont {C.~J.}\ \bibnamefont {Lobb}}, \bibinfo
  {author} {\bibfnamefont {J.~R.}\ \bibnamefont {Anderson}}, \bibinfo {author}
  {\bibfnamefont {F.~C.}\ \bibnamefont {Wellstood}}, \ and\ \bibinfo {author}
  {\bibfnamefont {E.}~\bibnamefont {Tiesinga}},\ }\href {\doibase
  10.1103/PhysRevB.77.214512} {\bibfield  {journal} {\bibinfo  {journal}
  {Physical Review B}\ }\textbf {\bibinfo {volume} {77}},\ \bibinfo {pages}
  {214512} (\bibinfo {year} {2008})},\ \bibinfo {note} {publisher: American
  Physical Society}\BibitemShut {NoStop}%
\bibitem [{\citenamefont {Iwasa}\ \emph {et~al.}(2019)\citenamefont {Iwasa},
  \citenamefont {Huang}, \citenamefont {Rai}, \citenamefont {Morosan},
  \citenamefont {Ohira-Kawamura},\ and\ \citenamefont
  {Nakajima}}]{iwasa_magnetic_2019}%
  \BibitemOpen
  \bibfield  {author} {\bibinfo {author} {\bibfnamefont {K.}~\bibnamefont
  {Iwasa}}, \bibinfo {author} {\bibfnamefont {C.-L.}\ \bibnamefont {Huang}},
  \bibinfo {author} {\bibfnamefont {B.~K.}\ \bibnamefont {Rai}}, \bibinfo
  {author} {\bibfnamefont {E.}~\bibnamefont {Morosan}}, \bibinfo {author}
  {\bibfnamefont {S.}~\bibnamefont {Ohira-Kawamura}}, \ and\ \bibinfo {author}
  {\bibfnamefont {K.}~\bibnamefont {Nakajima}},\ }\href
  {http://arxiv.org/abs/1810.11238} {\bibfield  {journal} {\bibinfo  {journal}
  {arXiv:1810.11238 [cond-mat]}\ } (\bibinfo {year} {2019})},\ \bibinfo {note}
  {arXiv: 1810.11238}\BibitemShut {NoStop}%
\bibitem [{\citenamefont {Fujii}\ \emph {et~al.}(2013)\citenamefont {Fujii},
  \citenamefont {Shibayama}, \citenamefont {Yamaguchi}, \citenamefont
  {Yoshida}, \citenamefont {Yamaguchi}, \citenamefont {Seki}, \citenamefont
  {Uchiyama}, \citenamefont {Baron},\ and\ \citenamefont
  {Umebayashi}}]{fujii_communication_2013}%
  \BibitemOpen
  \bibfield  {author} {\bibinfo {author} {\bibfnamefont {K.}~\bibnamefont
  {Fujii}}, \bibinfo {author} {\bibfnamefont {M.}~\bibnamefont {Shibayama}},
  \bibinfo {author} {\bibfnamefont {T.}~\bibnamefont {Yamaguchi}}, \bibinfo
  {author} {\bibfnamefont {K.}~\bibnamefont {Yoshida}}, \bibinfo {author}
  {\bibfnamefont {T.}~\bibnamefont {Yamaguchi}}, \bibinfo {author}
  {\bibfnamefont {S.}~\bibnamefont {Seki}}, \bibinfo {author} {\bibfnamefont
  {H.}~\bibnamefont {Uchiyama}}, \bibinfo {author} {\bibfnamefont {A.~Q.~R.}\
  \bibnamefont {Baron}}, \ and\ \bibinfo {author} {\bibfnamefont
  {Y.}~\bibnamefont {Umebayashi}},\ }\href {\doibase 10.1063/1.4802768}
  {\bibfield  {journal} {\bibinfo  {journal} {The Journal of Chemical Physics}\
  }\textbf {\bibinfo {volume} {138}},\ \bibinfo {pages} {151101} (\bibinfo
  {year} {2013})},\ \bibinfo {note} {publisher: American Institute of
  Physics}\BibitemShut {NoStop}%
\bibitem [{\citenamefont {Ghosh}\ and\ \citenamefont
  {Hasse}(1981)}]{ghosh_coherent_1981}%
  \BibitemOpen
  \bibfield  {author} {\bibinfo {author} {\bibfnamefont {G.}~\bibnamefont
  {Ghosh}}\ and\ \bibinfo {author} {\bibfnamefont {R.~W.}\ \bibnamefont
  {Hasse}},\ }\href {\doibase 10.1103/PhysRevA.24.1621} {\bibfield  {journal}
  {\bibinfo  {journal} {Physical Review A}\ }\textbf {\bibinfo {volume} {24}},\
  \bibinfo {pages} {1621} (\bibinfo {year} {1981})},\ \bibinfo {note}
  {publisher: American Physical Society}\BibitemShut {NoStop}%
\bibitem [{\citenamefont {Verriere}\ and\ \citenamefont
  {Mumpower}(2021)}]{verriere_improvements_2021}%
  \BibitemOpen
  \bibfield  {author} {\bibinfo {author} {\bibfnamefont {M.}~\bibnamefont
  {Verriere}}\ and\ \bibinfo {author} {\bibfnamefont {M.~R.}\ \bibnamefont
  {Mumpower}},\ }\href {\doibase 10.1103/PhysRevC.103.034617} {\bibfield
  {journal} {\bibinfo  {journal} {Physical Review C}\ }\textbf {\bibinfo
  {volume} {103}},\ \bibinfo {pages} {034617} (\bibinfo {year} {2021})},\
  \bibinfo {note} {publisher: American Physical Society}\BibitemShut {NoStop}%
\bibitem [{\citenamefont {Isar}\ and\ \citenamefont
  {Sandulescu}(2006)}]{isar_damped_2006}%
  \BibitemOpen
  \bibfield  {author} {\bibinfo {author} {\bibfnamefont {A.}~\bibnamefont
  {Isar}}\ and\ \bibinfo {author} {\bibfnamefont {A.}~\bibnamefont
  {Sandulescu}},\ }\href {http://arxiv.org/abs/quant-ph/0602149} {\bibfield
  {journal} {\bibinfo  {journal} {arXiv:quant-ph/0602149}\ } (\bibinfo {year}
  {2006})},\ \bibinfo {note} {arXiv: quant-ph/0602149}\BibitemShut {NoStop}%
\bibitem [{\citenamefont {Wu}\ \emph {et~al.}(1988)\citenamefont {Wu},
  \citenamefont {Li}, \citenamefont {Maruhn}, \citenamefont {Greiner},\ and\
  \citenamefont {Zhuo}}]{wu_quantum_1988}%
  \BibitemOpen
  \bibfield  {author} {\bibinfo {author} {\bibfnamefont {X.}~\bibnamefont
  {Wu}}, \bibinfo {author} {\bibfnamefont {Z.}~\bibnamefont {Li}}, \bibinfo
  {author} {\bibfnamefont {J.~A.}\ \bibnamefont {Maruhn}}, \bibinfo {author}
  {\bibfnamefont {W.}~\bibnamefont {Greiner}}, \ and\ \bibinfo {author}
  {\bibfnamefont {Y.}~\bibnamefont {Zhuo}},\ }\href {\doibase
  10.1088/0305-4616/14/8/008} {\bibfield  {journal} {\bibinfo  {journal}
  {Journal of Physics G: Nuclear Physics}\ }\textbf {\bibinfo {volume} {14}},\
  \bibinfo {pages} {1049} (\bibinfo {year} {1988})},\ \bibinfo {note}
  {publisher: IOP Publishing}\BibitemShut {NoStop}%
\bibitem [{\citenamefont {Um}\ \emph {et~al.}(1987)\citenamefont {Um},
  \citenamefont {Yeon},\ and\ \citenamefont {Kahng}}]{um_quantum_1987}%
  \BibitemOpen
  \bibfield  {author} {\bibinfo {author} {\bibfnamefont {C.~I.}\ \bibnamefont
  {Um}}, \bibinfo {author} {\bibfnamefont {K.~H.}\ \bibnamefont {Yeon}}, \ and\
  \bibinfo {author} {\bibfnamefont {W.~H.}\ \bibnamefont {Kahng}},\ }\href
  {\doibase 10.1088/0305-4470/20/3/024} {\bibfield  {journal} {\bibinfo
  {journal} {Journal of Physics A: Mathematical and General}\ }\textbf
  {\bibinfo {volume} {20}},\ \bibinfo {pages} {611} (\bibinfo {year} {1987})},\
  \bibinfo {note} {publisher: IOP Publishing}\BibitemShut {NoStop}%
\bibitem [{\citenamefont {Caldeira}\ and\ \citenamefont
  {Leggett}(1983)}]{caldeira_quantum_1983}%
  \BibitemOpen
  \bibfield  {author} {\bibinfo {author} {\bibfnamefont {A.~O.}\ \bibnamefont
  {Caldeira}}\ and\ \bibinfo {author} {\bibfnamefont {A.~J.}\ \bibnamefont
  {Leggett}},\ }\href {\doibase 10.1016/0003-4916(83)90202-6} {\bibfield
  {journal} {\bibinfo  {journal} {Annals of Physics}\ }\textbf {\bibinfo
  {volume} {149}},\ \bibinfo {pages} {374} (\bibinfo {year}
  {1983})}\BibitemShut {NoStop}%
\bibitem [{\citenamefont {Engels}(1975)}]{engels_helmholtz_1975}%
  \BibitemOpen
  \bibfield  {author} {\bibinfo {author} {\bibfnamefont {E.}~\bibnamefont
  {Engels}},\ }\href {\doibase 10.1007/BF02738572} {\bibfield  {journal}
  {\bibinfo  {journal} {Il Nuovo Cimento B Series 11}\ }\textbf {\bibinfo
  {volume} {26}},\ \bibinfo {pages} {481} (\bibinfo {year} {1975})}\BibitemShut
  {NoStop}%
\bibitem [{\citenamefont {Bateman}(1931)}]{bateman_dissipative_1931}%
  \BibitemOpen
  \bibfield  {author} {\bibinfo {author} {\bibfnamefont {H.}~\bibnamefont
  {Bateman}},\ }\href {\doibase 10.1103/PhysRev.38.815} {\bibfield  {journal}
  {\bibinfo  {journal} {Physical Review}\ }\textbf {\bibinfo {volume} {38}},\
  \bibinfo {pages} {815} (\bibinfo {year} {1931})},\ \bibinfo {note}
  {publisher: American Physical Society}\BibitemShut {NoStop}%
\bibitem [{\citenamefont {Caldirola}(1941)}]{caldirola_forze_1941}%
  \BibitemOpen
  \bibfield  {author} {\bibinfo {author} {\bibfnamefont {P.}~\bibnamefont
  {Caldirola}},\ }\href {\doibase 10.1007/BF02960144} {\bibfield  {journal}
  {\bibinfo  {journal} {Il Nuovo Cimento (1924-1942)}\ }\textbf {\bibinfo
  {volume} {18}},\ \bibinfo {pages} {393} (\bibinfo {year} {1941})}\BibitemShut
  {NoStop}%
\bibitem [{\citenamefont {Kanai}(1948)}]{kanai_quantization_1948}%
  \BibitemOpen
  \bibfield  {author} {\bibinfo {author} {\bibfnamefont {E.}~\bibnamefont
  {Kanai}},\ }\href {\doibase 10.1143/ptp/3.4.440} {\bibfield  {journal}
  {\bibinfo  {journal} {Progress of Theoretical Physics}\ }\textbf {\bibinfo
  {volume} {3}},\ \bibinfo {pages} {440} (\bibinfo {year} {1948})}\BibitemShut
  {NoStop}%
\bibitem [{\citenamefont {Feshbach}\ and\ \citenamefont
  {Tikochinsky}(1977)}]{feshbach_quantization_1977}%
  \BibitemOpen
  \bibfield  {author} {\bibinfo {author} {\bibfnamefont {H.}~\bibnamefont
  {Feshbach}}\ and\ \bibinfo {author} {\bibfnamefont {Y.}~\bibnamefont
  {Tikochinsky}},\ }\href {\doibase 10.1111/j.2164-0947.1977.tb02946.x}
  {\bibfield  {journal} {\bibinfo  {journal} {Transactions of the New York
  Academy of Sciences}\ }\textbf {\bibinfo {volume} {38}},\ \bibinfo {pages}
  {44} (\bibinfo {year} {1977})}\BibitemShut {NoStop}%
\bibitem [{\citenamefont {Banerjee}\ and\ \citenamefont
  {Mukherjee}(2002)}]{banerjee_canonical_2002}%
  \BibitemOpen
  \bibfield  {author} {\bibinfo {author} {\bibfnamefont {R.}~\bibnamefont
  {Banerjee}}\ and\ \bibinfo {author} {\bibfnamefont {P.}~\bibnamefont
  {Mukherjee}},\ }\href {\doibase 10.1088/0305-4470/35/27/305} {\bibfield
  {journal} {\bibinfo  {journal} {Journal of Physics A: Mathematical and
  General}\ }\textbf {\bibinfo {volume} {35}},\ \bibinfo {pages} {5591}
  (\bibinfo {year} {2002})},\ \bibinfo {note} {publisher: IOP
  Publishing}\BibitemShut {NoStop}%
\bibitem [{\citenamefont {Gitman}\ and\ \citenamefont
  {Kupriyanov}(2007)}]{gitman_action_2007}%
  \BibitemOpen
  \bibfield  {author} {\bibinfo {author} {\bibfnamefont {D.~M.}\ \bibnamefont
  {Gitman}}\ and\ \bibinfo {author} {\bibfnamefont {V.~G.}\ \bibnamefont
  {Kupriyanov}},\ }\href {\doibase 10.1088/1751-8113/40/33/010} {\bibfield
  {journal} {\bibinfo  {journal} {Journal of Physics A: Mathematical and
  Theoretical}\ }\textbf {\bibinfo {volume} {40}},\ \bibinfo {pages} {10071}
  (\bibinfo {year} {2007})},\ \bibinfo {note} {publisher: IOP
  Publishing}\BibitemShut {NoStop}%
\bibitem [{\citenamefont {Serhan}\ \emph {et~al.}(2018)\citenamefont {Serhan},
  \citenamefont {Abusini}, \citenamefont {Al-Jamel}, \citenamefont
  {El-Nasser},\ and\ \citenamefont {Rabei}}]{serhan_quantization_2018}%
  \BibitemOpen
  \bibfield  {author} {\bibinfo {author} {\bibfnamefont {M.}~\bibnamefont
  {Serhan}}, \bibinfo {author} {\bibfnamefont {M.}~\bibnamefont {Abusini}},
  \bibinfo {author} {\bibfnamefont {A.}~\bibnamefont {Al-Jamel}}, \bibinfo
  {author} {\bibfnamefont {H.}~\bibnamefont {El-Nasser}}, \ and\ \bibinfo
  {author} {\bibfnamefont {E.~M.}\ \bibnamefont {Rabei}},\ }\href {\doibase
  10.1063/1.5022321} {\bibfield  {journal} {\bibinfo  {journal} {Journal of
  Mathematical Physics}\ }\textbf {\bibinfo {volume} {59}},\ \bibinfo {pages}
  {082105} (\bibinfo {year} {2018})},\ \bibinfo {note} {publisher: American
  Institute of Physics}\BibitemShut {NoStop}%
\bibitem [{\citenamefont {Rabei}\ and\ \citenamefont
  {Al-Jamel}(2019)}]{rabei_quantisation_2019}%
  \BibitemOpen
  \bibfield  {author} {\bibinfo {author} {\bibfnamefont {E.~M.}\ \bibnamefont
  {Rabei}}\ and\ \bibinfo {author} {\bibfnamefont {A.}~\bibnamefont
  {Al-Jamel}},\ }\href {\doibase 10.1007/s12043-019-1882-4} {\bibfield
  {journal} {\bibinfo  {journal} {Pramana}\ }\textbf {\bibinfo {volume} {94}},\
  \bibinfo {pages} {1} (\bibinfo {year} {2019})}\BibitemShut {NoStop}%
\bibitem [{\citenamefont {Senitzky}(1960)}]{senitzky_dissipation_1960}%
  \BibitemOpen
  \bibfield  {author} {\bibinfo {author} {\bibfnamefont {I.~R.}\ \bibnamefont
  {Senitzky}},\ }\href {\doibase 10.1103/PhysRev.119.670} {\bibfield  {journal}
  {\bibinfo  {journal} {Physical Review}\ }\textbf {\bibinfo {volume} {119}},\
  \bibinfo {pages} {670} (\bibinfo {year} {1960})},\ \bibinfo {note}
  {publisher: American Physical Society}\BibitemShut {NoStop}%
\bibitem [{\citenamefont {Ford}\ \emph {et~al.}(1965)\citenamefont {Ford},
  \citenamefont {Kac},\ and\ \citenamefont {Mazur}}]{ford_statistical_1965}%
  \BibitemOpen
  \bibfield  {author} {\bibinfo {author} {\bibfnamefont {G.~W.}\ \bibnamefont
  {Ford}}, \bibinfo {author} {\bibfnamefont {M.}~\bibnamefont {Kac}}, \ and\
  \bibinfo {author} {\bibfnamefont {P.}~\bibnamefont {Mazur}},\ }\href
  {\doibase 10.1063/1.1704304} {\bibfield  {journal} {\bibinfo  {journal}
  {Journal of Mathematical Physics}\ }\textbf {\bibinfo {volume} {6}},\
  \bibinfo {pages} {504} (\bibinfo {year} {1965})},\ \bibinfo {note}
  {publisher: American Institute of Physics}\BibitemShut {NoStop}%
\bibitem [{\citenamefont {Dekker}(1977)}]{dekker_quantization_1977}%
  \BibitemOpen
  \bibfield  {author} {\bibinfo {author} {\bibfnamefont {H.}~\bibnamefont
  {Dekker}},\ }\href {\doibase 10.1103/PhysRevA.16.2126} {\bibfield  {journal}
  {\bibinfo  {journal} {Physical Review A}\ }\textbf {\bibinfo {volume} {16}},\
  \bibinfo {pages} {2126} (\bibinfo {year} {1977})},\ \bibinfo {note}
  {publisher: American Physical Society}\BibitemShut {NoStop}%
\bibitem [{\citenamefont {Thorwart}\ \emph {et~al.}(2004)\citenamefont
  {Thorwart}, \citenamefont {Paladino},\ and\ \citenamefont
  {Grifoni}}]{thorwart_dynamics_2004}%
  \BibitemOpen
  \bibfield  {author} {\bibinfo {author} {\bibfnamefont {M.}~\bibnamefont
  {Thorwart}}, \bibinfo {author} {\bibfnamefont {E.}~\bibnamefont {Paladino}},
  \ and\ \bibinfo {author} {\bibfnamefont {M.}~\bibnamefont {Grifoni}},\ }\href
  {\doibase 10.1016/j.chemphys.2003.10.007} {\bibfield  {journal} {\bibinfo
  {journal} {Chemical Physics}\ }\bibinfo {series} {The {Spin}-{Boson}
  {Problem}: {From} {Electron} {Transfer} to {Quantum} {Computing} ... to the
  60th {Birthday} of {Professor} {Ulrich} {Weiss}},\ \textbf {\bibinfo {volume}
  {296}},\ \bibinfo {pages} {333} (\bibinfo {year} {2004})}\BibitemShut
  {NoStop}%
\bibitem [{\citenamefont {Wilhelm}\ \emph {et~al.}(2004)\citenamefont
  {Wilhelm}, \citenamefont {Kleff},\ and\ \citenamefont {von
  Delft}}]{wilhelm_spin-boson_2004}%
  \BibitemOpen
  \bibfield  {author} {\bibinfo {author} {\bibfnamefont {F.~K.}\ \bibnamefont
  {Wilhelm}}, \bibinfo {author} {\bibfnamefont {S.}~\bibnamefont {Kleff}}, \
  and\ \bibinfo {author} {\bibfnamefont {J.}~\bibnamefont {von Delft}},\ }\href
  {\doibase 10.1016/j.chemphys.2003.10.010} {\bibfield  {journal} {\bibinfo
  {journal} {Chemical Physics}\ }\bibinfo {series} {The {Spin}-{Boson}
  {Problem}: {From} {Electron} {Transfer} to {Quantum} {Computing} ... to the
  60th {Birthday} of {Professor} {Ulrich} {Weiss}},\ \textbf {\bibinfo {volume}
  {296}},\ \bibinfo {pages} {345} (\bibinfo {year} {2004})}\BibitemShut
  {NoStop}%
\bibitem [{\citenamefont {Isar}\ \emph {et~al.}(1993)\citenamefont {Isar},
  \citenamefont {Sandulescu},\ and\ \citenamefont
  {Scheid}}]{isar_density_1993}%
  \BibitemOpen
  \bibfield  {author} {\bibinfo {author} {\bibfnamefont {A.}~\bibnamefont
  {Isar}}, \bibinfo {author} {\bibfnamefont {A.}~\bibnamefont {Sandulescu}}, \
  and\ \bibinfo {author} {\bibfnamefont {W.}~\bibnamefont {Scheid}},\ }\href
  {\doibase 10.1063/1.530013} {\bibfield  {journal} {\bibinfo  {journal}
  {Journal of Mathematical Physics}\ }\textbf {\bibinfo {volume} {34}},\
  \bibinfo {pages} {3887} (\bibinfo {year} {1993})},\ \bibinfo {note}
  {publisher: American Institute of Physics}\BibitemShut {NoStop}%
\bibitem [{\citenamefont {Isar}(1999)}]{isar_uncertainty_1999}%
  \BibitemOpen
  \bibfield  {author} {\bibinfo {author} {\bibfnamefont {A.}~\bibnamefont
  {Isar}},\ }\href {\doibase
  10.1002/(SICI)1521-3978(199909)47:7/8<855::AID-PROP855>3.0.CO;2-Z} {\bibfield
   {journal} {\bibinfo  {journal} {Fortschritte der Physik}\ }\textbf {\bibinfo
  {volume} {47}},\ \bibinfo {pages} {855} (\bibinfo {year} {1999})}\BibitemShut
  {NoStop}%
\bibitem [{\citenamefont {Fujii}(2013)}]{fujii_quantum_2013}%
  \BibitemOpen
  \bibfield  {author} {\bibinfo {author} {\bibfnamefont {K.}~\bibnamefont
  {Fujii}},\ }\href {\doibase 10.5772/52671} {\emph {\bibinfo {title} {Quantum
  Damped Harmonic Oscillator}}}\ (\bibinfo  {publisher} {IntechOpen},\ \bibinfo
  {year} {2013})\ \bibinfo {note} {publication Title: Advances in Quantum
  Mechanics}\BibitemShut {NoStop}%
\bibitem [{\citenamefont {Greenberger}(1979)}]{greenberger_critique_1979}%
  \BibitemOpen
  \bibfield  {author} {\bibinfo {author} {\bibfnamefont {D.~M.}\ \bibnamefont
  {Greenberger}},\ }\href {\doibase 10.1063/1.524148} {\bibfield  {journal}
  {\bibinfo  {journal} {Journal of Mathematical Physics}\ }\textbf {\bibinfo
  {volume} {20}},\ \bibinfo {pages} {762} (\bibinfo {year} {1979})},\ \bibinfo
  {note} {publisher: American Institute of Physics}\BibitemShut {NoStop}%
\bibitem [{\citenamefont {Hasse}(1975)}]{hasse_quantum_1975}%
  \BibitemOpen
  \bibfield  {author} {\bibinfo {author} {\bibfnamefont {R.~W.}\ \bibnamefont
  {Hasse}},\ }\href {\doibase 10.1063/1.522431} {\bibfield  {journal} {\bibinfo
   {journal} {Journal of Mathematical Physics}\ }\textbf {\bibinfo {volume}
  {16}},\ \bibinfo {pages} {2005} (\bibinfo {year} {1975})},\ \bibinfo {note}
  {publisher: American Institute of Physics}\BibitemShut {NoStop}%
\bibitem [{\citenamefont {Bagarello}\ \emph
  {et~al.}(2019{\natexlab{a}})\citenamefont {Bagarello}, \citenamefont
  {Gargano},\ and\ \citenamefont {Roccati}}]{bagarello_no-go_2019}%
  \BibitemOpen
  \bibfield  {author} {\bibinfo {author} {\bibfnamefont {F.}~\bibnamefont
  {Bagarello}}, \bibinfo {author} {\bibfnamefont {F.}~\bibnamefont {Gargano}},
  \ and\ \bibinfo {author} {\bibfnamefont {F.}~\bibnamefont {Roccati}},\ }\href
  {\doibase 10.1016/j.physleta.2019.06.022} {\bibfield  {journal} {\bibinfo
  {journal} {Physics Letters A}\ }\textbf {\bibinfo {volume} {383}},\ \bibinfo
  {pages} {2836} (\bibinfo {year} {2019}{\natexlab{a}})}\BibitemShut {NoStop}%
\bibitem [{\citenamefont {Deguchi}\ and\ \citenamefont
  {Fujiwara}(2019)}]{deguchi_square-integrable_2019}%
  \BibitemOpen
  \bibfield  {author} {\bibinfo {author} {\bibfnamefont {S.}~\bibnamefont
  {Deguchi}}\ and\ \bibinfo {author} {\bibfnamefont {Y.}~\bibnamefont
  {Fujiwara}},\ }\href {http://arxiv.org/abs/1910.08271} {\bibfield  {journal}
  {\bibinfo  {journal} {arXiv:1910.08271 [hep-th, physics:math-ph,
  physics:quant-ph]}\ } (\bibinfo {year} {2019})},\ \bibinfo {note} {arXiv:
  1910.08271}\BibitemShut {NoStop}%
\bibitem [{\citenamefont {Bagarello}\ \emph
  {et~al.}(2019{\natexlab{b}})\citenamefont {Bagarello}, \citenamefont
  {Gargano},\ and\ \citenamefont {Roccati}}]{bagarello_reply_2019}%
  \BibitemOpen
  \bibfield  {author} {\bibinfo {author} {\bibfnamefont {F.}~\bibnamefont
  {Bagarello}}, \bibinfo {author} {\bibfnamefont {F.}~\bibnamefont {Gargano}},
  \ and\ \bibinfo {author} {\bibfnamefont {F.}~\bibnamefont {Roccati}},\ }\href
  {http://arxiv.org/abs/1910.12561} {\bibfield  {journal} {\bibinfo  {journal}
  {arXiv:1910.12561 [math-ph]}\ } (\bibinfo {year} {2019}{\natexlab{b}})},\
  \bibinfo {note} {arXiv: 1910.12561}\BibitemShut {NoStop}%
\bibitem [{\citenamefont {Bagarello}\ \emph {et~al.}(2020)\citenamefont
  {Bagarello}, \citenamefont {Gargano},\ and\ \citenamefont
  {Roccati}}]{bagarello_remarks_2020}%
  \BibitemOpen
  \bibfield  {author} {\bibinfo {author} {\bibfnamefont {F.}~\bibnamefont
  {Bagarello}}, \bibinfo {author} {\bibfnamefont {F.}~\bibnamefont {Gargano}},
  \ and\ \bibinfo {author} {\bibfnamefont {F.}~\bibnamefont {Roccati}},\ }\href
  {\doibase 10.1016/j.aop.2020.168091} {\bibfield  {journal} {\bibinfo
  {journal} {Annals of Physics}\ }\textbf {\bibinfo {volume} {414}},\ \bibinfo
  {pages} {168091} (\bibinfo {year} {2020})}\BibitemShut {NoStop}%
\bibitem [{\citenamefont {Deguchi}\ and\ \citenamefont
  {Fujiwara}(2020)}]{deguchi_quantization_2020}%
  \BibitemOpen
  \bibfield  {author} {\bibinfo {author} {\bibfnamefont {S.}~\bibnamefont
  {Deguchi}}\ and\ \bibinfo {author} {\bibfnamefont {Y.}~\bibnamefont
  {Fujiwara}},\ }\href {\doibase 10.1103/PhysRevA.101.022105} {\bibfield
  {journal} {\bibinfo  {journal} {Physical Review A}\ }\textbf {\bibinfo
  {volume} {101}},\ \bibinfo {pages} {022105} (\bibinfo {year} {2020})},\
  \bibinfo {note} {publisher: American Physical Society}\BibitemShut {NoStop}%
\bibitem [{\citenamefont {Griffiths}(2005)}]{griffiths_introduction_2005}%
  \BibitemOpen
  \bibfield  {author} {\bibinfo {author} {\bibfnamefont {D.~J.}\ \bibnamefont
  {Griffiths}},\ }\href@noop {} {\emph {\bibinfo {title} {Introduction to
  {Quantum} {Mechanics}}}},\ \bibinfo {edition} {2nd}\ ed.\ (\bibinfo
  {publisher} {Pearson Education},\ \bibinfo {address} {New Jersey},\ \bibinfo
  {year} {2005})\BibitemShut {NoStop}%
\bibitem [{Note1()}]{Note1}%
  \BibitemOpen
  \bibinfo {note} {As justified earlier, here we set $\tau _0 =
  0$.}\BibitemShut {Stop}%
\end{thebibliography}%

\end{document}